\documentclass[%
 reprint,
superscriptaddress,
 amsmath,amssymb,
 aps,
]{revtex4-1}

\usepackage{graphicx}
\usepackage{dcolumn}
\usepackage{bm}

\begin{document}

\preprint{APS/123-QED}

\title{Probing Strongly Correlated 4$f$-Orbital Symmetry of the Ground 
State \\ in Yb Compounds by Linear Dichroism in Core-Level Photoemission}

\author{Takeo Mori}
\author{Satoshi Kitayama}
\author{Yuina Kanai}
\author{Sho Naimen}
\author{Hidenori Fujiwara}
\affiliation{Division of Materials Physics, Graduate School of Engineering Science, Osaka University, Toyonaka, Osaka 560-8531, Japan}

\author{Atsushi Higashiya}
\affiliation{Faculty of Science and Engineering, Setsunan University, Neyagawa, Osaka 572-8508, Japan }
\affiliation{RIKEN SPring-8 Center, Sayo, Hyogo 679-5148, Japan}

\author{Kenji Tamasaku}
\affiliation{RIKEN SPring-8 Center, Sayo, Hyogo 679-5148, Japan}
\affiliation{Center for Promotion of Advanced and Interdisciplinary Research, Graduate School of 
Engineering Science, Osaka University, Toyonaka, Osaka 560-8531, Japan}

\author{Arata Tanaka}
\affiliation{Department of Quantum Matter, ADSM, Hiroshima University, Higashi-Hiroshima, Hiroshima 739-8530, Japan}

\author{Kensei Terashima}
 \altaffiliation[Present address: ]{Research Laboratory for Surface Science, Okayama University, Okayama 700-8530, Japan}
\author{Shin Imada}
\affiliation{Department of Physical Science, Ritsumeikan University, Kusatsu, Shiga 525-8577, Japan }

\author{Akira Yasui}
\affiliation{Japan Synchrotron Radiation Research Institute, SPring-8, Sayo, Hyogo 679-5198, Japan}

\author{Yuji Saitoh}
\affiliation{Condensed Matter Science Division, Japan Atomic Energy Agency, SPring-8, Hyogo 679-5148, Japan}

\author{Kohei Yamagami}
\author{Kohei Yano}
\author{Taiki Matsumoto}
\affiliation{Division of Materials Physics, Graduate School of Engineering Science, Osaka University, Toyonaka, Osaka 560-8531, Japan}

\author{Takayuki Kiss}
\affiliation{Division of Materials Physics, Graduate School of Engineering Science, Osaka University, Toyonaka, Osaka 560-8531, Japan}
\affiliation{RIKEN SPring-8 Center, Sayo, Hyogo 679-5148, Japan}
\affiliation{Center for Promotion of Advanced and Interdisciplinary Research, Graduate School of 
Engineering Science, Osaka University, Toyonaka, Osaka 560-8531, Japan}

\author{Makina Yabashi} 
\author{Tetsuya Ishikawa}
\affiliation{RIKEN SPring-8 Center, Sayo, Hyogo 679-5148, Japan}

\author{Shigemasa Suga}
\affiliation{RIKEN SPring-8 Center, Sayo, Hyogo 679-5148, Japan}
\affiliation{Institute of Scientific and Industrial Research, Osaka University, Ibaraki, Osaka 567-0047, Japan}

\author{Yoshichika $\bar{\rm O}$nuki}
 \altaffiliation[Present address: ]{Faculty of Science, University of the Ryukyus, Nishihara, Okinawa 903-0213, Japan}
\affiliation{Department of Physics, Graduate School of Science, Osaka University, Toyonaka, Osaka 560-0043, Japan}

\author{Takao Ebihara}
\affiliation{Department of Physics, Shizuoka University, Shizuoka 422-8529, Japan}

\author{Akira Sekiyama}
 \email{sekiyama@mp.es.osaka-u.ac.jp}
\affiliation{Division of Materials Physics, Graduate School of Engineering Science, Osaka University, Toyonaka, Osaka 560-8531, Japan}
\affiliation{RIKEN SPring-8 Center, Sayo, Hyogo 679-5148, Japan}
\affiliation{Center for Promotion of Advanced and Interdisciplinary Research, Graduate School of 
Engineering Science, Osaka University, Toyonaka, Osaka 560-8531, Japan}

\date{\today}

\begin{abstract}
We show that the strongly correlated $4f$-orbital symmetry 
of the ground state  
is revealed by linear dichroism in core-level photoemission spectra 
as we have discovered for YbRh$_2$Si$_2$ and YbCu$_2$Si$_2$. 
Theoretical analysis tells us that the linear dichroism reflects the anisotropic 
charge distributions resulting from crystalline electric field. 
We have successfully determined the ground-state $4f$ symmetry  
for both compounds from the polarization-dependent {\it angle-resolved} core-level spectra 
at a low temperature well below the first excitation energy. 
The excited-state symmetry is also probed by temperature dependence of 
the linear dichroism where the high measuring temperatures 
are of the order of the crystal-field-splitting energies.
\end{abstract}

\maketitle


\section{Intrroduction}

\begin{figure}
\begin{center}
\includegraphics[width=8cm]{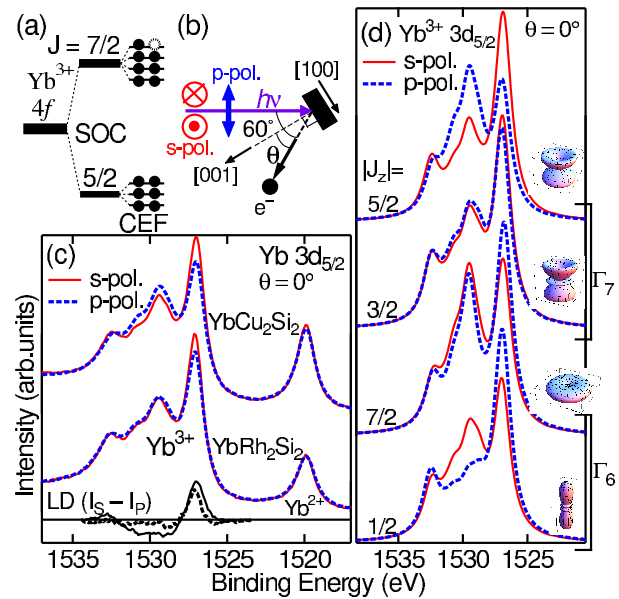}
\end{center}
\vspace{5mm}
\caption{\label{bothLD}(Color online) (a) Schematically drawn Yb$^{3+} 4f$ levels split by the spin-orbit 
coupling (SOC) and further split by the crystalline electric field (CEF) in tetragonal symmetry. 
Filled (open) circle denotes an occupied 4$f$ electron (hole).
(b) Geometry for the LD in HAXPES measurements, where $\theta$ is the angle 
of photoelectron detection direction to the [001] direction (normal direction to the 
cleaved surfaces).  
(c) Polarization-dependent Yb 3$d_{5/2}$ core-level HAXPES spectra  
for YbRh$_2$Si$_2$ and YbCu$_2$Si$_2$ at $\theta =0^\circ$ 
(along the [001] direction). The spectra are normalized by the overall $3d_{5/2}$ 
spectral weight displayed in this graph. 
LD for YbRh$_2$Si$_2$ (YbCu$_2$Si$_2$) is also shown by the dashed (solid) 
line in the lower panel.
(d) Simulated polarization-dependent 3$d_{5/2}$ photoemission spectra 
for the Yb$^{3+}$ ions assuming the pure $J_z$ ground state, 
together with the corresponding 4$f$-hole spatial distributions.}
\end{figure}

Strongly correlated electron systems show a variety of intriguing phenomena like 
unconventional and/or high-temperature superconductivity, spin and charge ordering, 
formation of heavy fermions, non-trivial (Kondo) semiconducting behavior, 
and quantum criticality. 
Among them, such Yb-based single-crystalline compounds as YbRh$_2$Si$_2$~\cite{YRS00} 
and $\beta$-YbAlB$_4$~\cite{YbAlB4SN08} showing the quantum criticality in ambient pressure, 
a Kondo semiconductor YbB$_{12}$~\cite{Iga98,Susaki99}, 
valence-fluctuating YbAl$_3$~\cite{YbAl3C02,YbAl3E03} and 
YbCu$_2$Si$_2$~\cite{RC2S2,YCS09},
and a very heavy fermionic YbCo$_2$Zn$_{20}$~\cite{PNASCo,SaigaCo,OnukiCo} 
have been synthesized with excellent quality 
and thus intensively studied within a couple of decades. 
Since the strong Coulomb repulsion (effective value of 6-10 eV)  works 
between the 4$f$ electrons in the Yb sites, an ionic picture is a good starting point to 
discuss and reveal their electronic structure as well as the origins of various phenomena 
in the crystalline solids. 
The majority of the Yb ions in the above-mentioned materials is in trivalent 4$f^{13}$ (one 4$f$ hole) 
configurations although there are to some extent Yb$^{2+}$ (4$f^{14}$) components due to the 
hybridization between the 4f orbitals and other valence-band states crossing the Fermi level. 
The Yb$^{3+}$ $4f$ levels are split by the spin-orbit coupling ($>1$ eV) and further split by the 
crystalline electric field (CEF, from a few to several tens meV) in solids 
as shown in Fig. \ref{bothLD}(a). 

Ground-state $f$-orbital symmetry determined by the CEF splitting 
is very fundamental information of the realistic strongly correlated 
electron systems. 
In contrast to the case of transition-metal oxides in which the electron 
correlations work among the $d$-orbital electrons, 
the ground-state symmetry is not straightforward revealed 
since it is unclear which sites act as effective {\it "ligands"} 
for the $f$ sites. 
A standard experimental technique for determining the 4$f$ levels with their 
symmetry is to analyze the inelastic neutron scattering spectra 
and anisotropy in the magnetic susceptibility of single crystals. 
However, the magnetic 4$f$-4$f$ excitations are often hampered 
by the phonon excitations with the same energy scale. 
Moreover, it is difficult to uniquely determine the symmetry of all $f$ levels 
by the analysis of the magnetic anisotropy since there are too many 
free parameters for a unique description of the CEF potential. 
Actually, the ground-state 4$f$-orbital symmetry is not clear 
for here reported YbRh$_2$Si$_2$ and YbCu$_2$Si$_2$ 
although a couple of possible solutions have been proposed. 
Polarized neutron scattering~\cite{WillersCePt3Si} for large single crystals is principally powerful to 
determine the $f$-orbital symmetry, but time-consuming. 

Recently linear dichroism (LD) in the 3$d$-to-4$f$ soft X-ray absorption 
spectroscopy (XAS)~\cite{WillersCePt3Si,Hansmann08LD,WillersCeTIn5,WillersCe122} 
for single crystals has been reported for the heavy fermion systems with nearly Ce$^{3+}$ 
($4f^1$) configurations as a powerful tool to determine the 4$f$ ground state 
owing to the dipole selection rules. 
However, it is difficult to apply this technique to probe the Yb$^{3+}$ states 
since there is only single-peak structure ($3d^94f^{14}$ final state) 
at the $M_5$ absorption edge. 
On the other hand, the selection rules work also in the photoemission process 
while the excited electron energy is much higher than that in the absorption process. 
We have discovered that the atomic-like multiplet-split structure 
in the core-level photoemission spectra shows easily detectable LD 
reflecting the outer 4$f$ spatial distribution probed by the created core hole. 
In this letter, we demonstrate that the 4$f$-orbital symmetry of the ground state 
for YbRh$_2$Si$_2$ and YbCu$_2$Si$_2$ in tetragonal symmetry 
is determined by the LD in the Yb 3$d_{5/2}$ core-level photoemission, 
and further that the excited-state $4f$ symmetry is also probed 
by the temperature-dependent data. 
YbRh$_2$Si$_2$ is known as the first discovered Yb system showing quantum 
criticality in ambient pressure as mentioned above. 
YbCu$_2$Si$_2$ is a counterpart of the first discovered heavy fermion superconductor 
CeCu$_2$Si$_2$~\cite{CeCuSi2} with respect to an electron-hole symmetry. 
However, it does not show superconductivity, being recognized as 
one of the valence-fluctuating systems with anisotropic magnetic 
susceptibility~\cite{RC2S2,YCS09,YCSPES} 
which implies the anisotropic 4$f$-hole spatial distribution. 

\section{Experimental}
Since the binding energy of the Yb 3$d_{5/2}$ core-level is higher than 1.5 keV, 
hard X-ray photoemission spectroscopy (HAXPES) at least $h\nu > 2.5$ keV is preferable 
to avoid the surface contributions deviated from the bulk ones~\cite{SugaYbAl3,KitayamaTokimeki}. 
We have performed LD in HAXPES~\cite{SekiyamaAuAg,ASHAXPES2013} 
at BL19LXU of SPring-8~\cite{YabashiPRL01}  
by using a MBS A1-HE hemispherical photoelectron spectrometer. 
A Si(111) double-crystal monochromator selected 7.9 keV radiation 
with linear polarization along the horizontal direction 
(the so-called degree of linear polarization $P_L > +0.98$), 
which was further monochromatized by a Si(620) channel-cut crystal. 
In order to switch the linear polarization of the excitation light from the horizontal 
to vertical directions, two single-crystalline (100) diamonds were used as a phase retarder 
with the (220) reflection placed downstream of the channel-cut crystal. 
$P_L$ of the polarization-switched x-ray after the phase retarder was estimated as 
$-0.96$, corresponding to the vertically linear polarization components of 98\%.  
Since the detection direction of 
photoelectrons was set in the horizontal plane 
with the angle to the incident photons of 60$^{\circ}$ as shown in Fig.~\ref{bothLD}(b),  
the experimental configuration 
at the horizontally (vertically) polarized light excitation corresponds to the p-polarization 
(s-polarization). 
The excitation light was focused onto the samples with the  spot size of 
$\sim$25 $\mu$m $\times$ 25 $\mu$m by using an ellipsoidal Kirkpatrik-Baez mirror.
The single crystals of YbRh$_2$Si$_2$ and YbCu$_2$Si$_2$ 
synthesized by a flux method were cleaved along the (001) plane {\it in situ}, 
where the base pressure was $\sim$1$\times10^{-7}$ Pa. 
The sample and surface quality was checked by the absence of 
any core-level spectral weight caused by a possible impurity including oxygen and carbon. 
The energy resolution was set to 250 meV. 

\section{Results and Discussions}
The polarization-dependent Yb 3$d_{5/2}$ core-level HAXPES spectra of 
YbRh$_2$Si$_2$ and YbCu$_2$Si$_2$ at $\theta = 0^{\circ}$ 
(photoelectron detection is along the [001] direction) 
are shown in Fig.\ref{bothLD}(c). 
There are a single peak at the binding energy of $\sim$1520 eV and 
a multiple-peak structure ranging from 1525 to 1535 eV in all spectra. 
Since the 4$f$ subshell is fully occupied in the Yb$^{2+}$ sites 
with spherically symmetric 4$f$ distribution, 
the former single peak is ascribed to the Yb$^{2+}$ states. 
The $3d^94f^{13}$ (one $4f$ hole with one $3d$ core hole) final states 
for the Yb$^{3+}$ components show the atomic-like multiplet-split peak structure 
in the 1525-1535 eV range. 
Clear LD defined by the difference of the spectral weight between the s- and p-polarization 
configurations is seen in the Yb$^{3+}$ $3d_{5/2}$ spectral weight depending on material, 
where it is relatively weaker for YbRh$_2$Si$_2$. 
For instance, 
one of the Yb$^{3+}$ $3d_{5/2}$ peak at 1527 eV is stronger in the s-polarization 
configuration (s-pol.) than in the p-polarization configuration (p-pol.) 
whereas a structure with the 1529.5-eV peak and 1530.5-eV shoulder is stronger 
in the p-pol. for both compounds. 
Possible photoelectron diffraction effects are ruled out of the origin of LD 
based on the fact that the degree of LD is mutually different between both compounds 
with the same ThCr$_2$Si$_2$ crystal structure. 

To clarify the origin of LD in the Yb$^{3+}$ $3d_{5/2}$ core-level HAXPES spectra, 
we have performed ionic calculations including the full multiplet 
theory~\cite{Thole85} and the local CEF splitting using the XTLS 9.0 program~\cite{XTLS}. 
All atomic parameters such as the 4$f$-4$f$ and 3$d$-4$f$ Coulomb and exchange 
interactions (Slater integrals) and the 3$d$ and 4$f$ spin-orbit couplings have been 
obtained by Cowan's code~\cite{Cowan} based on the Hartree-Fock method. 
The Slater integrals (spin-orbit couplings) are reduced down to 88\% (98\%) to fit 
the core-level photoemission spectra~\cite{YbB12PRB2009}. 
We show the polarization-dependent core-level spectra at $\theta=0^{\circ}$ 
for pure $|J_z\rangle$ states of the Yb$^{3+}$ ions with $J = 7/2$ in Fig.\ref{bothLD}(d). 
LD depends strongly on $|J_z|$, 
reflecting the Coulomb interactions between the 3$d$ and 4$f$ holes 
with anisotropic spatial distributions 
in the final state. 
The result of the simulations tells us that 
the observed LD in the core-level HAXPES spectra originates from the 
anisotropic 4$f$ hole distribution under CEF in the initial state. 

In the case of Yb$^{3+}$ ions in tetragonal symmetry, the eightfold degenerate 
$J=7/2$ state splits into four doublets as 
\begin{eqnarray}
&&|\Gamma_7^1\rangle = c|\pm5/2\rangle+\sqrt{1-c^2}|\mp3/2\rangle,\label{G71} \\
&&|\Gamma_7^2\rangle =- \sqrt{1-c^2}|\pm5/2\rangle+c|\mp3/2\rangle,\label{G72} \\
&&|\Gamma_6^1\rangle = b|\pm1/2\rangle+\sqrt{1-b^2}|\mp7/2\rangle,\label{G61}\\
&&|\Gamma_6^2\rangle =\sqrt{1-b^2}|\pm1/2\rangle-b|\mp7/2\rangle,\label{G62}
\end{eqnarray}
where the coefficients $0\leq b\leq 1, 0\leq c\leq 1$ defining the actual charge distributions, 
and CEF splitting energies depend on the CEF parameters 
$B_2^0, B_4^0, B_4^4, B_6^0$, and $B_6^4$ in Stevens formalism~\cite{Stevens}. 
Since all CEF splitting energies are highly expected to be much larger ($\gtrsim100$ K) than 
the measured temperature of 14 K for YbRh$_2$Si$_2$~\cite{YRSINS1} 
and YbCu$_2$Si$_2$~\cite{INS82,JETP98}, 
it is justified to assume that only the lowest state is populated.   
As shown in Fig.\ref{bothLD}(d), LD is qualitatively different between $|\Gamma_6^{1,2}\rangle$ 
and $|\Gamma_7^{1,2}\rangle$, where LDs for $|J_z|=5/2$ and 3/2 are qualitatively 
consistent with those for the experimental data. 
Therefore, the $|\Gamma_6^{1,2}\rangle$ ground state formed by the $|J_z|$ = 7/2 and 
1/2 components is ruled out  
for both compounds. 

\begin{figure}
\includegraphics[width=7.5cm]{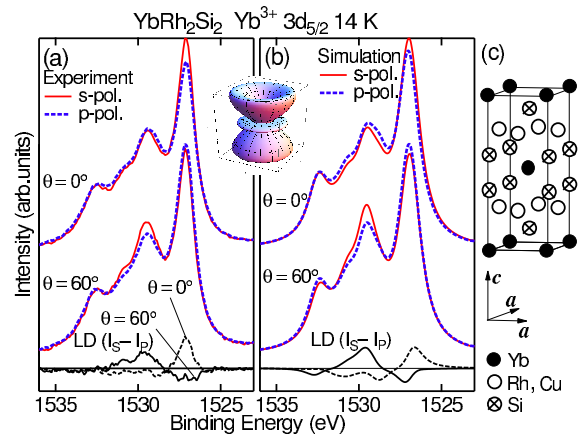}
\vspace{2mm}
\caption{(Color online) (a) Polarization-dependent Yb$^{3+}$ $3d_{5/2}$ core-level HAXPES spectra 
of YbRh$_2$Si$_2$ at $\theta = 0^\circ$ and 60$^\circ$ and their LDs, 
where the Shirley-type background has been subtracted from the raw spectra. 
The spectra are normalized by the Yb$^{3+}$ $3d_{5/2}$ spectral weight.
(b) Simulated polarization-dependent core-level photoemission spectra and their LDs 
[dashed (solid) line for $\theta=0^\circ$ (60$^\circ$)]
for the Yb$^{3+}$ ion with the $|J_z|=3/2$ ($\Gamma_7$) ground state 
at the same geometrical configurations as those for the experiments. 
The inset shows the corresponding 4$f$ hole spatial distribution in the initial state.
(c) Crystal structure of YbRh$_2$Si$_2$ and YbCu$_2$Si$_2$.}
\label{YRSLD}
\end{figure}

To more accurately determine the ground-state orbital symmetry, we have also performed 
the polarization-dependent core-level HAXPES for YbRh$_2$Si$_2$ at different $\theta$ 
of 60$^\circ$ corresponding to the incident photon direction parallel to the [001] direction 
as shown in Fig.\ref{YRSLD}(a). 
Compared to LD at $\theta=0^\circ$, the sign of LD at $\theta=60^\circ$ is flipped 
as recognized at the bottom of the figure. 
We have found that our data are best described by the pure $|J_z| = 3/2$ ground state 
as shown in Fig.\ref{YRSLD}(b).  
So far, it has been unclear whether $|\Gamma_6^1 \rangle$ with dominant $|J_z|=1/2$ 
or $|\Gamma_7^1 \rangle$ with dominant $|J_z|=3/2$ forms the ground 
state~\cite{YRSCEF1,YRSCEF2}
while a comparison of the slab calculations for subsurface YbRh$_2$Si$_2$ to 
the low-energy angle-resolved photoemission data has suggested the 
$|\Gamma_7^1 \rangle$ ground state~\cite{YRSARPES10}. 
Here the $|J_z|=3/2$ ($\Gamma_7$) ground state is unambiguously revealed 
for YbRh$_2$Si$_2$ from our LD in HAXPES at $\theta=0^\circ$ and 60$^\circ$.
Since there is no anisotropy within the $ab$ plane for the 4$f$ hole distribution 
with the $|J_z|=3/2$ ground state as depicted in the inset of Fig. \ref{YRSLD}, 
it is naturally concluded that the Yb 4$f$ holes are hybridized with the partially filled neighbor 
Rh 4$d$ and Si 3$sp$ states where the $4f$ hole distribution spreads over both sites 
in Fig.\ref{YRSLD}(c).


\begin{figure}
\begin{center}
\includegraphics[width=7cm]{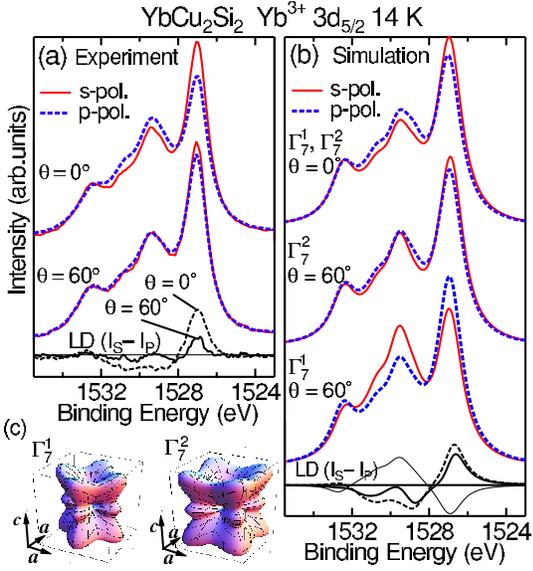}
\end{center}
\vspace{2mm}
\begin{center}
\caption{(Color online) (a) Polarization-dependent Yb$^{3+}$ $3d_{5/2}$ core-level HAXPES spectra 
of YbCu$_2$Si$_2$ at $\theta = 0^\circ$ and 60$^\circ$ and their LDs, 
where the Shirley-type background has been subtracted from the raw spectra. 
(b) Simulated polarization-dependent core-level photoemission spectra and their LDs 
at the same geometrical configurations as those for the experiments 
for the Yb$^{3+}$ ion with the $|\Gamma_7^1\rangle$ and 
$|\Gamma_7^2\rangle$ ground states with the 
$|J_z|=3/2$ (5/2) component of 87\% (13\%). 
LD at $\theta=0^\circ$ is represented by a dashed line whereas that at 
$\theta=60^\circ$ for the $|\Gamma_7^1\rangle$ ($|\Gamma_7^2\rangle$) ground state 
is shown by a thin (thick) solid line.
(c) 4$f$ hole spatial distribution for the initial $|\Gamma_7^1\rangle$ and 
$|\Gamma_7^2\rangle$ states.}
\label{YCSLD}
\end{center}
\end{figure}

The polarization-dependent Yb$^{3+}$ $3d_{5/2}$ HAXPES spectra of YbCu$_2$Si$_2$ 
at $\theta=0^\circ$ and 60$^\circ$ are shown in Fig.\ref{YCSLD}(a). 
In contrast to the data for YbRh$_2$Si$_2$, the sign of LD is not flipped 
between two angles of $\theta$ whereas LD is reduced at  $\theta=60^\circ$, 
being inconsistent with the simulations for the pure $|J_z| = 3/2$ state in Fig.\ref{YRSLD}(b). 
Our detailed analysis indicates that the data set of polarization-dependent HAXPES 
is best described by the ground state of 
\begin{equation}
|\Gamma_7^2\rangle = -0.36|\pm5/2\rangle+0.93|\mp3/2\rangle \label{YCSGS} 
\end{equation}
as shown in Fig.\ref{YCSLD}(b), where the precision of $c^2$ in Eq.(\ref{G72}) is $\pm$0.05.  
The state of $|\Gamma_7^1\rangle = 0.36|\pm5/2\rangle+0.93|\mp3/2\rangle$ 
with the same amount of the $|J_z|=3/2$ components as in Eq.(\ref{YCSGS}), 
of which the spatial 4$f$ hole distribution shows the same shape as for $|\Gamma_7^2\rangle$ 
with rotation within the $ab$ plane by 45$^\circ$ [see Fig.\ref{YCSLD}(c), hereafter called 
as in-plane rotation], gives the same 
spectra and LD as those for $|\Gamma_7^2\rangle$ at $\theta=0^\circ$. 
Therefore, the data at $\theta=60^\circ$ 
enable us to discriminate the in-plane rotation 
of the $4f$ charge distributions 
and unambiguously determine the ground state. 
The 4$f$ hole distribution for $|\Gamma_7^2\rangle$ is elongated along the Si sites 
not the Cu sites with fully occupied 3$d$ levels~\cite{YCS09}
as shown in Figs.\ref{YRSLD}(c) and \ref{YCSLD}(c), 
leading to the conclusion that the 4$f$ holes are primarily hybridized with 
the Si 3$sp$ states. 

The polarization-dependent Yb$3d_{5/2}$ core-level HAXPES spectra 
shown here are well reproduced apart from the Yb$^{2+}$ contributions 
by the simulations for the atomic-like models 
as seen in the analysis of many LD data in XAS 
for Ce compounds~\cite{WillersCePt3Si,Hansmann08LD,WillersCeTIn5,WillersCe122}, 
where the hybridization effects are not explicitly taken into account. 
Such a successful analysis 
is owing to the highly localized nature of the Yb$^{3+}$ sites in the 
$3d$ core-level photoemission final states due to the core hole-$4f$ Coulomb 
interactions ($\sim$10 eV) giving a sufficient energy splitting between 
the Yb$^{3+}$ and Yb$^{2+}$ final states, 
and the configuration dependence of the hybridization strengths~\cite{Cdep} 
leading to the reduced hybridization 
in the final states. 
The analysis needs to be extended by using the Anderson impurity model 
for strongly hybridized systems showing the core-level spectral line shape 
highly deviated from the atomic-like multplet-split structure, 
which is not the case for the data displayed here.

\begin{figure}
\begin{center}
\includegraphics[width=8.5cm]{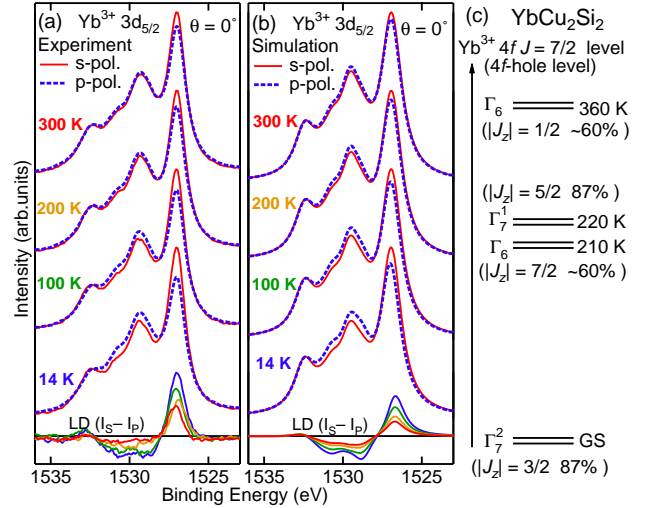}
\end{center}
\vspace{2mm}
\begin{center}
\caption{(Color online)
(a) Temperature dependence of the polarization-dependent Yb$^{3+}$ $3d_{5/2}$ HAXPES 
spectra at $\theta=0^\circ$ and their LDs for YbCu$_2$Si$_2$.
(b) Simulated temperature dependence of the polarization-dependent 
core-level photoemission spectra and their LDs at $\theta=0^\circ$ 
for the Yb$^{3+}$ ion with the 4$f$ levels in (c), 
where the coefficient $b^2$ in Eqs. (\ref{G61}) and (\ref{G62}) 
for the excited $|\Gamma_6^{1,2}\rangle$ states of 0.4 
with the lower state with predominant $|J_z|=7/2$ components. 
(c) Schematically drawn Yb$^{3+} 4f$ $J=7/2$ levels with symmetry 
split by CEF determined for YbCu$_2$Si$_2$ together with the fraction 
of the predominant component for each doublet. 
}
\label{LDTdep}
\end{center}
\end{figure}

We have also performed the temperature- and polarization-dependent HAXPES 
at $\theta=0^\circ$ for YbCu$_2$Si$_2$ as shown in Fig.\ref{LDTdep}(a), 
verifying that LD is reduced at high temperatures without a flip of its sign. 
The temperature dependence originates from a partial occupation of 
the excited state $i$ split by $\Delta_i$ from the ground state 
at high temperatures $T$  
with a fraction of $\exp[-\Delta_i/(k_BT)]/\{1+\sum_i\exp[-\Delta_i/(k_BT)]\}$, 
which leads to the isotropic spectra 
at a sufficiently high temperature. 
Taking the fact that the magnetic excitations are clearly seen 
at $\sim$210 and $\sim$360K for YbCu$_2$Si$_2$~\cite{INS82} into account, 
we can determine to some extent the $4f$-orbital symmetry of the excited states 
based on our data and simulations, 
where $|\Gamma_7^1\rangle$ with dominant $|J_z|=5/2$ gives 
larger LD than the experimental one and $|\Gamma_6\rangle$ formed by the 
the $|J_z|=7/2$ and 1/2 states shows a sign-flipped LD compared to the data 
as suggested by the simulations in Fig. \ref{bothLD}(d). 
Figure \ref{LDTdep}(b) shows one of the best simulated 
temperature-dependent HAXPES spectra and LDs 
for the $4f$-level scheme with symmetry shown in Fig. \ref{LDTdep}(c), 
which has been optimized with some ambiguities 
in determining the parameters. 
The other $4f$-level scheme with the first excitation energy of $\sim$100 K
similar to a previously proposed one~\cite{JETP98} 
is inconsistent with the experimental result 
since the simulations with this scheme 
give larger temperature dependence of LD involving a sign-flipping 
or an enhancement of LD at 100 K. 
%
Here optimized CEF splitting energies for the $4f$-level scheme in Fig.\ref{LDTdep}(c) 
is similar to another previously proposed one in Ref.~\onlinecite{RC2S2}  
based on the magnetic excitations in Ref.~\onlinecite{INS82}. 


\section{Conclusions}
In conclusion, we have shown that the 4$f$-orbital symmetry of the ground  states 
as well as that of the excited states in the Yb compounds 
is probed by the LD in the core-level HAXPES and its $\theta$ dependence. 
The ground-state symmetry is unambiguously determined by the LD 
under the sole assumption that only the lowest state is populated. 
LD in the core-level HAXPES has the advantage over LD in XAS 
in discriminating the same shape of the 4$f$ charge distributions with in-plane rotation 
as shown for YbCu$_2$Si$_2$ owing to the experimental parameter $\theta$ 
in addition to the polarization. 
The discrimination of the in-plane rotation of charge distribution is also feasible 
by the polarization-dependent non-resonant inelastic X-ray scattering~\cite{NIXSCeCu2Si2}, 
but the throughput and energy resolution are much better for LD in the core-level HAXPES. 
Therefore, the experimental technique demonstrated here will be very promising 
to reveal the strongly correlated orbital symmetry of the ground and excited states 
in the atomic-like partially filled subshell in solids, complementing the neutron scattering.

\begin{acknowledgments}
We thank T. Kadono, F. Honda, Y. Nakata, Y. Nakatani, T. Yamaguchi, H. Fuchimoto, 
T. Yagi, S. Tachibana, A. Yamasaki, and Y. Tanaka for supporting the experiments. 
This work was supported by Grant-in-Aid for Scientific Research (23654121), 
that for Young Scientists (23684027,23740240,25800205), 
that for Innovative Areas (20102003), 
the Global COE (G10) from MEXT and JSPS, Japan, 
and by Toray Science Foundation. 
The hard x-ray photoemission was performed at SPring-8 under the approval of 
JASRI (2014A1149). 
\end{acknowledgments}


\end{document}